\documentstyle[prl,aps,epsf]{revtex}
\begin{document} \draft 
%\preprint{Presented in the XVI International Conference on Thermoelectrics, Dresden, Germany (1997).}
% The following line is crucial to get two-column format right
\twocolumn[\hsize\textwidth\columnwidth\hsize\csname @twocolumnfalse\endcsname
\title{Geometric contribution to the measurement 
of thermoelectric power and Nernst 
coefficient in a strong magnetic field}
\author{Kazuaki IKEDA, Hiroaki Nakamura$^a$, and
Satarou Yamaguchi$^a$}
\address{Department of Fusion Science, The Graduate University for 
Advanced Studies,
322-6, Oroshi-Cho, Toki-City, Gifu-Prefecture, 509-52, Japan, Phone\&Fax : +81-52-789-4538, e-mail:ikeda@rouge.nifs.ac.jp}
\address{$^a$National Institute for Fusion Science, Oroshi-Cho, Toki-City, Gifu-Prefecture, 509-52, Japan}
\date{\mbox{This paper was presented in the XVI International Conference on Thermoelectrics, Dresden, Germany (1997).} } 
\maketitle

\begin{abstract}
On the measurement of thermoelectric power and Nernst 
coefficient, we used two kinds of shapes for a sample. One is 
``Bridge shape" and the other is, we call,``Fat-Bridge shape". 
The latter has 5 times wider main body than the former. We 
used pure n-InSb in this experiment, whose carrier condensation 
measured at 77K was $6.6 \times 10^{-14} {\rm cm}^{-3}.$
The length of sample is 17mm and temperature difference 
induced between the edges in 
that direction were about $10^{\circ}$C or $100^{\circ}$C near a room 
temperature range of 0 to $100^{\circ}$C. Magnetic induction applied in 
the perpendicular direction to temperature gradient was in the 
range of 0 up to 4 Tesla.  In the case of ``Fat-Bridge shape", we 
detected about 10\%  smaller Nernst coefficient and 1 to 10\% 
smaller thermoelectric power comparing to the ``Bridge shape". 
We suppose this phenomena is due to the geometric 
contribution on the different shape of samples.
\end{abstract}
%\pacs{75.10.Jm}
\vskip2pc] \narrowtext

\section{Introduction}
We call some semiconductors Nernst element~\cite{ref1,ref2}, which will 
be able to use for the generation of electric power by applying 
the Nernst effect. As the fundamental study for the power 
generation by the Nernst elements, we are studying for the 
transport properties of its candidates in a magnetic field~\cite{ref3,ref4}. 
By measuring their transport coefficients, we will be able to 
estimate the efficiency of energy conversion.

\section{Experiment}
The thermoelectric effect and the Nernst effect can be written 
respectively in the form,
\begin{eqnarray}
{\mbox {\boldmath $E$} } &=& \alpha \cdot {\mbox{\it grad }} T ,  \label{eq.1} \\
{\mbox {\boldmath $E$} } &=& N \cdot {\mbox {\boldmath $B$} }\times{\mbox{\it grad }}T , \label{eq.2}
\end{eqnarray}
where {\boldmath $E$} is electric field, $\alpha$ thermoelectric power, $T$ 
temperature, $N$ Nernst coefficient and {\boldmath $B$} magnetic induction.

\begin{figure}
\epsfxsize=8cm \epsffile{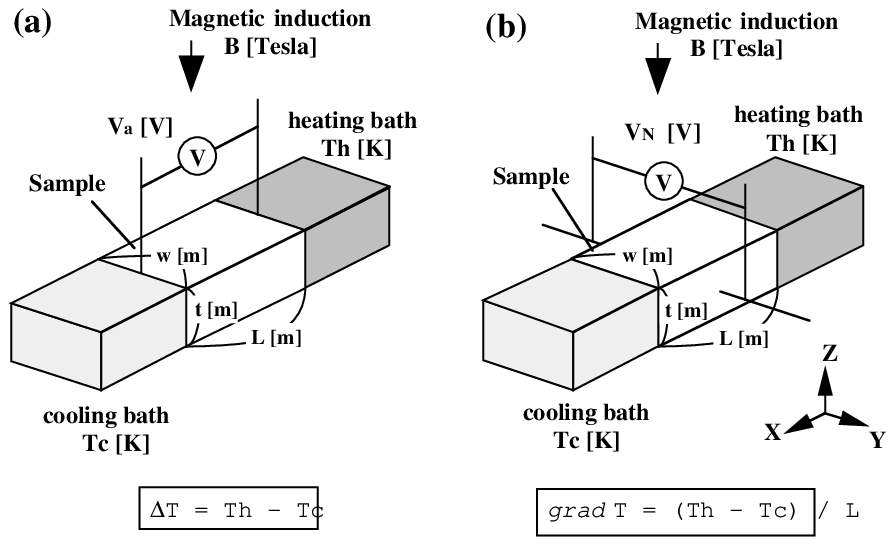}
%\epsfile{width=8cm,file=Fig1.eps} 
\caption{The scheme of measurement. (a) is for thermoelectric
power  and (b) is for Nernst coefficient.}
\label{fig.1}
\end{figure}

Suppose a sample is rectangular parallelepiped and its scale is 
of length $L,$ width $w,$ thickness $t,$ then a temperature gradient is 
given in the direction of $L.$ That scheme is indicated to FIG.~\ref{fig.1}.
 
We can obtain the thermoelectric power $\alpha$ by detecting the 
potential difference $V_{\alpha}$ and temperature difference $\delta T$ between 
edges in the direction of $L,$ and the Nernst coefficient $N$ by 
temperature gradient $\Delta T/L$ and the magnetic induction $B$ applied 
in the perpendicular direction to temperature gradient and 
potential difference $V_{\rm N}$ in the perpendicular direction to  both 
temperature gradient and magnetic induction. They were 
calculated from eqs.(\ref{eq.1}) and (\ref{eq.2}) as follows:
\begin{eqnarray}
V_{\rm \alpha} &=& \alpha \cdot 
\left( \frac{\Delta T}{L} \right) , \label{eq.3}\\
V_{\rm N} &=& w N B \cdot 
\left( \frac{\Delta T}{L} \right) . \label{eq.4}
\end{eqnarray}
Using eqs.(\ref{eq.3}) and (\ref{eq.4}), 
we obtain the following relation:
\begin{eqnarray}
\alpha &=& 
\left( \frac{V_{\rm \alpha} \cdot L}{\Delta T}  \right) , \label{eq.3.1}\\
N &=& 
\left( \frac{V_{\rm N} }{B } \right) 
\left( \frac{L }{w } \right) . \label{eq.4.1}
\end{eqnarray}
%%%%%%%%%%%%%%%%%%%%

%%%%%%%%%%%%%%%%%%%%
On this measurement,  it is very important that we used  two 
kinds of the shapes for a sample. They are indicated to FIG.~\ref{fig.2}.
%%%%%%%%%%%%%%%%%%%%
One is called ``Bridge shape", that has narrow main body, 
several legs for measuring leads and wide heads on the edges for 
having good attachments to heating or cooling bath. We call 
the other  ``Fat-Bridge shape", that has 5 times wider body than 
Bridge one. The samples were cut out from thin wafers at 
accuracy of about 0.1mm by wire cutter. 
We measured their 
scale at precision of within 0.005mm by micrometer.
These samples were located on the sample holder as is indicated 
to FIG.~\ref{fig.3}. 
%%%%%%%%%%%%%%%%%%%%%%%%%%%%%%%%%%%%%
\begin{figure}
\epsfxsize=8cm \epsffile{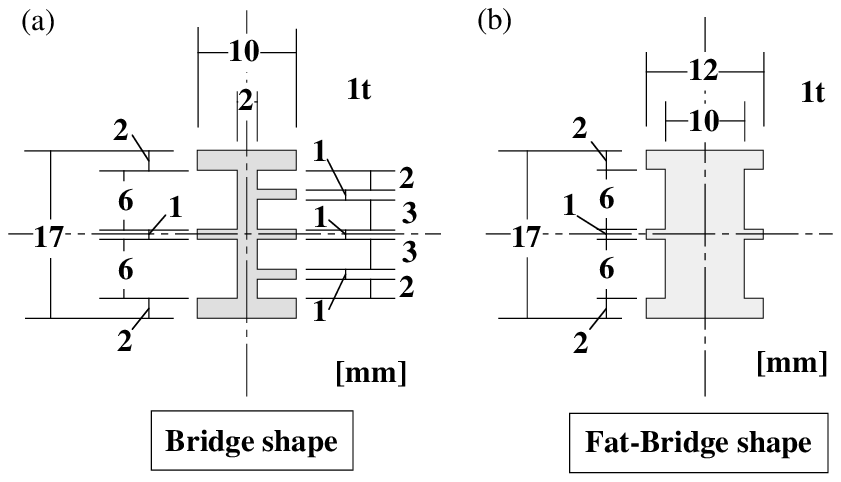}
%\epsfile{width=8cm,file=Fig2.eps} 
\caption{The shape of samples. (a) is Bridge shape and (b) is
Fat-Bridge shape.}
\label{fig.2}
\end{figure}

%%%%%%%%%%%%%%%%%%%%%%%%%%%%%%%%%%%%%
\begin{figure}
\epsfxsize=8cm \epsffile{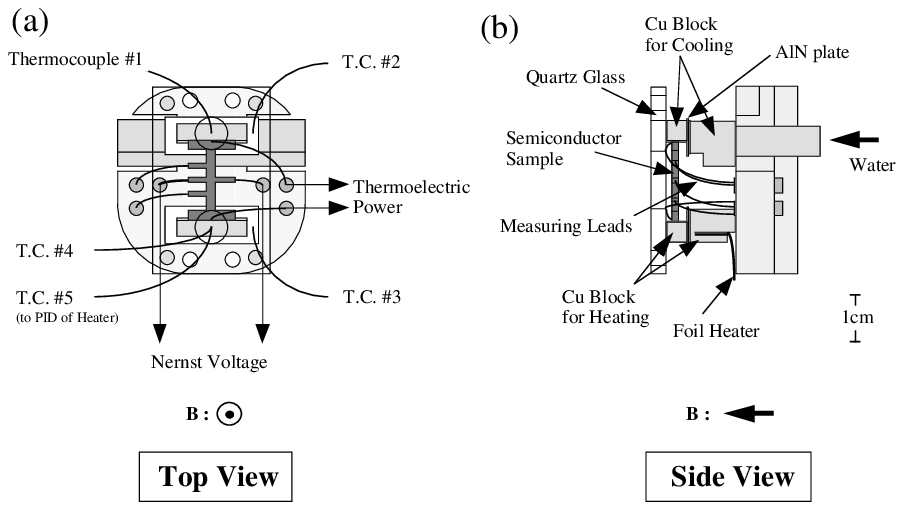}
%\epsfile{width=8cm,file=Fig3.eps} 
\caption{Sample holder which is mounted on sample mounter 
and inserted in the central part of vacuum chamber. 
Figures (a) and (b) are its top view and side view respectively.}
\label{fig.3}
\end{figure}

When we make the temperature gradient in the samples, 
temperatures are controlled by two Cu blocks attached to the 
edges of sample. One is heated up to about $100^{\circ}$C by foil heater 
and the other is cooled down to about $0^{\circ}$C by the unfrozen 
liquid mixed water and ethylene glycol. Temperatures were 
measured by Chromel-Almuel thermocouples. Measuring leads 
for voltage signals are 0.5mm diameter wires of the Cu which 
were welded to sample.
We had the experiment around room temperatures and on the 
two kinds of temperature conditions. On one condition, the 
temperature difference of between sample edges is $10^{\circ}$C, and 
temperature of cold side is increased 0 up to $80^{\circ}$C by $10^{\circ}$C. 
On the other condition, the cold side is fixed to $0^{\circ}$C and hot 
side is $100^{\circ}$C.
Experimental equipments are indicated to FIG.~\ref{fig.4}. By using the 
superconducting magnet coil which is included into the 
cryostat, magnetic field of 0 to 4 Tesla is generated in the 
central region of vacuum chamber. The region of 20mm cube 
in the central part of the chamber  is stable within 0.2\% to the 
magnetic field strength of central point. We had measurements 
of physical properties in that region. Inner pressure of chamber 
was less than $10^{-3}$Pa.
%%%%%%%%%%%%%
\begin{figure}
\epsfxsize=8cm \epsffile{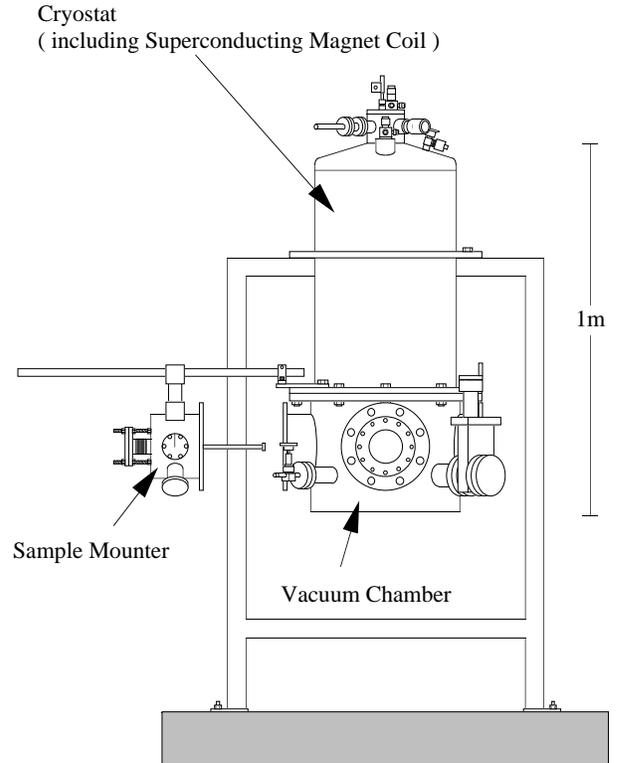}
%\epsfile{width=8cm,file=Fig4.eps} 
\caption{The experimental equipments. Superconducting magnet 
           coil, vacuum chamber and sample mounter.}
\label{fig.4}
\end{figure}

Physical phenomena of temperature and electric field are 
transduced to voltage signals by thermocouples and leads. As is 
indicated to FIG.~\ref{fig.5}, transduced voltage signals are inputted to 
isolation amplifiers. Then, amplified signals are sent to 16bit 
A/D converter plugged into a personal computer and are 
digitized. 
The maximum gain of the Amplifier is 2000. We can detect 
voltage signals in the resolution of 0.15$\mu$V at maximum. On
this measurement, the precision of the temperature 
measurement were less than 0.1K  and the relative error of the 
voltage measurement were not exceed 0.5\%.
%%%%%%%%%%%%%
\begin{figure}
\epsfxsize=8cm \epsffile{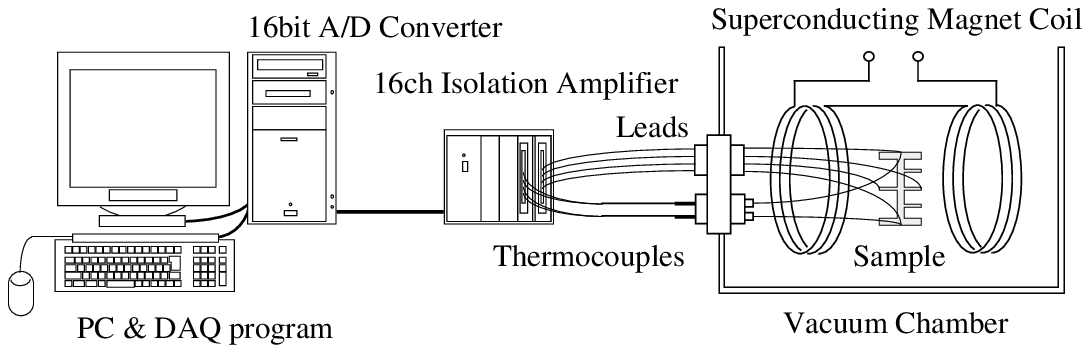}
%\epsfile{width=8cm,file=Fig5.eps} 
\caption{Data acquisition system.}
\label{fig.5}
\end{figure}

\section{Data analysis and discussion}
Figures \ref{fig.6} and \ref{fig.7} show the dependence of thermoelectric power 
and Nernst coefficient on magnetic induction respectively. They 
include data on the conditions that the temperature pairs of 
heating block and cooling one are set on 0 and $100^{\circ}$C, 10 and 
$20^{\circ}$C and 80 and $90^{\circ}$C. In FIG.~\ref{fig.6}, we used
\begin{equation}
\beta = N \cdot B , \label{eq.5}
\end{equation}
as substitute for Nernst coefficient.

%%%%%%%%%%%%%
\begin{figure}
\epsfxsize=8cm \epsffile{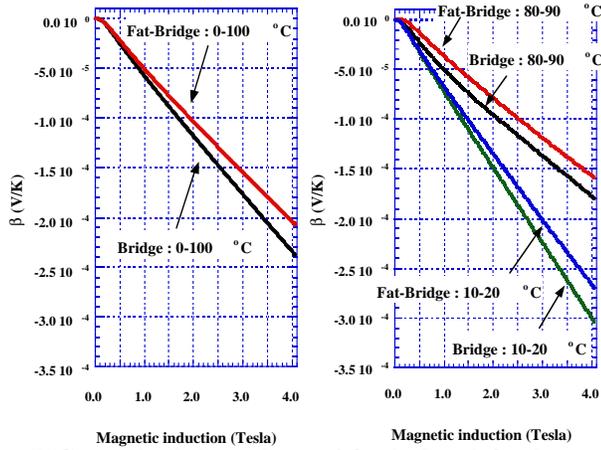}
%\epsfile{width=8cm,file=Fig6.eps} 
\caption{The B dependence of $\beta$, which is defined as product of
Nernst coefficient and magnetic induction.}
\label{fig.6}
\end{figure}

In very weak and strong magnetic fields, $\beta$ was linearly 
decreasing or decreasing with the increase of field strength. It 
means that Nernst coefficient is nearly constant in their 
regions. In intermediate fields, there is a transition of $N$ from 
one constant value to another. Especially in the case of 
$80-90^{\circ}$C, the sign of coefficient changed. Except in weak field 
of the case of $80-90^{\circ}$C, the sign of Nernst coefficients was 
negative on this experiment. These results have qualitative 
agreements with the investigation of up to 2 Tesla reported in
Ref.~\cite{ref5}. But we detected the qualitative difference when the shapes 
of samples were different. At 4 Tesla, Nernst coefficients 
measured on the Fat-Bridge shape were about 12\% smaller than 
those on the Bridge shape in all cases and that tendencies are 
similar in strong fields. 
In the measurement of thermoelectric powers, 
the any values on 
the Fat-Bridge shape was smaller than that on the Bridge shape. 
But the rate of its reduction were about 8\%, 5\% and less than 
1\% in the case of 
$0-100^{\circ}$C, $10-20^{\circ}$C, 
$80-90^{\circ}$C respectively. 

We suppose that the detected differences are apparently and due 
to the difference of samples' shape. We call it geometric 
contribution. To interpret this contribution and the inner state 
of the samples, more detailed analysis and the theoretical 
investigation will be proceeded in near future.
Figure \ref{fig.8} shows the dependence of thermoelectric power on the 
applied direction of magnetic induction, which obtained on the 
measurement of ``Fat Bridge shape" in $\Delta T = 100^{\circ}$C.
The sign of magnetic induction appears the applied direction of that. 
Figure~\ref{fig.8}
includes the measured data, the analytically derived component 
and the analytically excluded one. Measured data was not 
symmetric to the direction of magnetic induction. We suppose 
this is due to the reason that the positions of measuring leads 
had the difference in the perpendicular direction to heat flux and 
Nernst effect was detected slightly. That component is indicated 
as the excluded $\beta$ in FIG.~\ref{fig.8}. Because of  the similar mechanism, 
thermoelectric effect was detected on the measurement of Nernst 
effect. That appeared as offset voltages on the measurement 
apparently. They are indicated in the FIG.~\ref{fig.9}, which obtained on 
the measurement of ``Fat Bridge shape" in $\Delta T = 100^{\circ}$C.
These  phenomena are detected on all our measurements.

%%%%%%%%%%%%%
\begin{figure}
\epsfxsize=8cm \epsffile{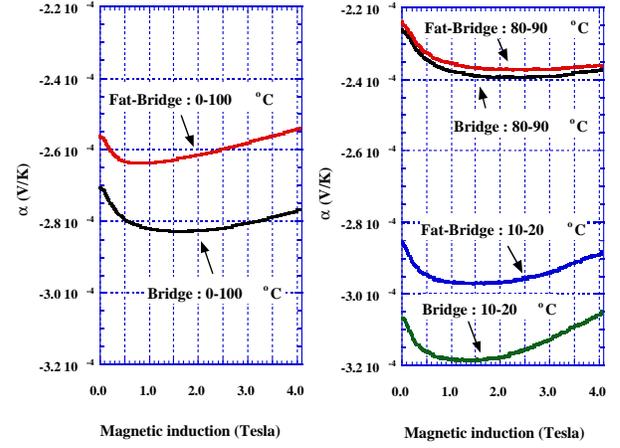}
%\epsfile{width=8cm,file=Fig7.eps} 
\caption{The B dependence of the thermoelectric power.}
\label{fig.7}
\end{figure}

%%%%%%%%%%%%%
\begin{figure}
\epsfxsize=8cm \epsffile{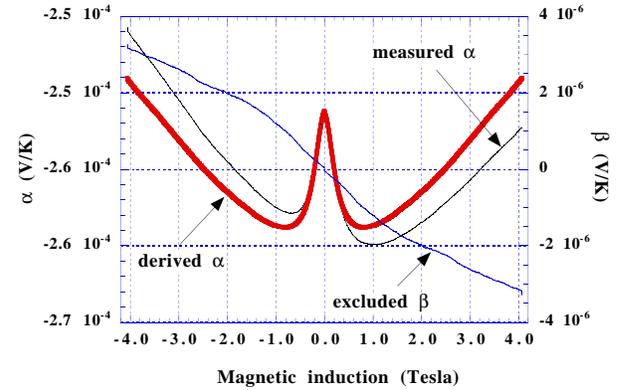}
%\epsfile{width=8cm,file=Fig8.eps} 
\caption{Measured $\alpha$, analytically derived $\alpha$ and 
excluded $\beta$.}
\label{fig.8}
\end{figure}
%%%%%%%%%%%%%
\begin{figure}
\epsfxsize=8cm \epsffile{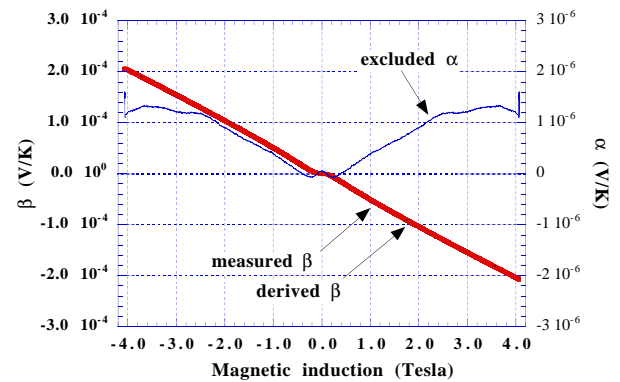}
%\epsfile{width=8cm,file=Fig9.eps} 
\caption{Measured $\beta$, analytically derived $\beta$ and 
excluded $\alpha.$}
\label{fig.9}
\end{figure}

To excluded the contaminated components on the measured 
data, we calculate $\alpha$ and $\beta$ as follows:
\begin{eqnarray}
\alpha&=& \frac{1}{2} 
 \left\{
    \left( \frac{V_{\rm \alpha} (B) }{\Delta T } \right)
  + \left( \frac{V_{\rm \alpha} (-B) }{\Delta T } \right)
 \right\}, 	\label{eq.6} \\
\beta&=& \frac{1}{2} 
 \left\{
    \left( \frac{V_{\rm  N} (B) }{\Delta T } \right)
  - \left( \frac{V_{\rm  N} (-B) }{\Delta T } \right)
 \right\}. 	\label{eq.7} 
\end{eqnarray}

The propriety of these processes are due to  the properties that 
thermoelectric effect is even to the direction of magnetic 
induction and Nernst effect is odd.

\section{Conclusion}
When the magnetic field is applied to the semiconductor in 
which heat flux exist, two components of electric fields are 
generated inside of it. One is in the origin of thermoelectric 
effect and the other is Nernst effect. 
The mixture of these two 
effects generate the geometric contributions on the 
measurement of physical properties as we detected on this 
measurement using ``Bridge shape" and ``Fat-Bridge shape". 
This condition is similar to the measurement 
of the Hall effect~\cite{ref6,ref7}. 
When we confirm the physical properties by the measurement, 
the geometric contributions must be considered. We shall need 
to develop the calculation code to interpret the transport 
properties including the heat and current fluxes in more details 
and self-consistently.

\acknowledgments
\noindent
The authors are grateful to Mr. Nishimura in Nishimura factory 
corporation for the process of samples and Dr. Tatsumi in 
Sumitomo Electric Industries for providing semiconductors and 
Prof. Kuroda in Nagoya university for many useful supports. 
We appreciate Prof. Iiyoshi and Prof. Motojima in the National 
Institute for Fusion Science for their helpful comments.

%\begin{center}
%{\bf References}
%\end{center}
%\noindent


\begin{references}
\bibitem{ref1} S. Yamaguchi, A. Iiyoshi, O. Motojima, M. Okamoto, 
     S. Sudo, M. Ohnishi, M. Onozuka and C. Uesono, {\it ``Direct
     Energy Conversion of Radiation Energy in Fusion Energy",
     Proc. of 7th Int. Conf. Merging Nucl. Energy Systems}
     (ICENES), (1994) 502.
\bibitem{ref2} S. Yamaguchi, K. Ikeda, H. Nakamura and K. Kuroda, {\it ``A 
     Nuclear Fusion Study of Thermoelectric Conversion in 
     Magnetic field", 4th Int. Sympo. on Fusion Nuclear Tech.}, 
     ND-P25, Tokyo, Japan, April (1997).
\bibitem{ref3} K. Ikeda, H. Nakamura, S. Yamaguchi and K. Kuroda,  
      J. Adv. Sci., {\bf 8}, 147 (1996), (in Japanese).
\bibitem{ref4} H. Nakamura,  K. Ikeda, S. Yamaguchi and K. Kuroda,  
      J. Adv. Sci., {\bf 8}, 153 (1996), (in Japanese).
\bibitem{ref5} Ya. Agaev, O. Mosanov, and O. Ismailov, Sov. Phys.
      Semicond., {\bf 1}, 711 (1967).
\bibitem{ref6}  H. Welker and  H. Weiss, Z. Phys. {\bf 138}, 322 (1954).
\bibitem{ref7}  H. Weiss,{\it  "Semiconductors and Semimetals 1", Academic,}
      New York (1966).
\end{references}
\end{document}